# Sistem Informasi Pengarsipan Menggunakan Algoritma Levensthein String pada Kecamatan Seberang Ulu II

Pratiwi Vidyarsih[1], Leon Andretti Abdillah[2], Ari Muzakir[3]
[1,2] Information Systems, Bina Darma University
[3] Informatics Engineering, Bina Darma University
Jalan Ahmad Yani No.3, Plaju, Palembang
[1] viddhye@gmail.com, [2] leon.abdillah@yahoo.com

**Abstract.** Archival information systems in government agency is one of the most used applications for daily acitivities. One feature in application management information document is searching. This feature serves to search for documents from a collection of available information based on keywords entered by the user. But some researches on a search engine (searching) concluded that the average user error in the search is quite high due to several factors. Therefore, we need a development on this feature as search suggestion. This study discusses the application of the method of approximate string matching algorithm using levenshtein distance. Levensthein distance algorithm is capable of calculating the minimum distance conversion of a string into another string to the optimum. Archiving information system using Levensthein Algorithm String is an application that will be built to address these problems, this application will help, especially in the administration to enter or save a document, locate and make a report that will be seen by government agencies.

**Kyewords:** Archival information systems, Levensthein algorithms.

## 1  Pendahuluan

Perkembangan teknologi dibidang informasi mendorong setiap instansi atau perusahaan untuk tetap mengikuti perkembangannya, terutama berkenaan dengan perkembangan teknologi informasi (TI) yang ada hubungannya dengan kegiatan perusahaan tersebut. Perkembangan TI telah memberikan kontribusi yang cukup berarti dalam meningkatkan kegiatan usaha khususnya dalam hal pengolahan data. Kemajuan tersebut meliputi, dapat dilihat tidak hanya dari dimensi perangkat keras (*hardware*) tetapi juga dari dimensi perangkat lunak (*software*), serta perkembangan kualitas sumber daya manusianya (*brainware*) [1]. Pelayanan sistem informasi (SI) memungkinkan pemakai mengakses data dan informasi lingkungan berdasarkan subsistem fungsional dan menggantikan teknologi atau sistem penyimpanan data konvensional ke dalam bentuk data yang dapat disimpan dalam komputer sehingga meningkatkan efisiensi dalam pencarian data dan perawatan data.

Arsip merupakan medium yang digunakan oleh tiap institusi [2]. Sistem Informasi Arsip adalah suatu sistem informasi yang mengelola data yang menyangkut pengumpulan, pengelolaan, pemusnahan, pencetakan laporan dan pencarian kembali arsip yang berbasis komputer sehingga mampu mengelola arsip dengan lebih efektif dan efesien dan pada akhirnya dapat memberi masukan informasi secara aktual dan akurat tentang perumusan kebijakan, strategi dan program pembangunan. Proses pengolahan arsip tersebut membutuhkan algoritma tertentu. Algoritma [3] merupakan langkah-langkah yang digunakan untuk menyelesaikan suatu permasalahan dengan pendekatan secara matematis. Algoritma yang digunakan pada penelitian ini adalah Algoritma Levensthein.

Algoritma Levensthein atau sering disebut dengan *Levensthein Distance* atau *Edit Distance* merupakan suatu algoritma *string* matriks untuk mengukur perbedaan antar string yang berbeda. Algoritma yang ditemukan oleh Vladimir Levenshtein, seorang ilmuwan Rusia, pada tahun 1965 [4]. Algoritma ini berguna untuk memeriksa kemiripan dari dua buah *string* yang umumnya ditemukan pada aplikasi-aplikasi pengecekan suatu ejaan. Ada 3 (tiga) macam operasi utama yang dapat dilakukan oleh algoritma Levensthein yaitu : 1) Pengubahan karakter, 2) Penambahan karakter, dan 3) Penghapusan karakter. Algoritma ini berjalan mulai dari pojok kiri atas sebuah array dua dimensi [5] yang telah diisi sejumlah karakter *string* awal dan *string* target dan diberikan nilai *cost*. Nilai *cost* pada ujung kanan bawah menjadi nilai *edit distance* yang menggambarkan jumlah perbedaan dua *string*.

Kecamatan Seberang Ulu II merupakan instansi pemerintahan yang melayani masyarakat seperti, administrasi surat menyurat berupa surat keterangan KTP, kartu keluarga, surat pengantar ke Walikota, surat datang dan pindahnya penduduk, surat kematian kependudukan, surat kegiatan pelayanan pencatatan akta-akta penting, surat pengantar untuk MTQ dan surat pengantar penduduk untuk gotong royong. Kecamatan Seberang Ulu II dalam menjalankan aktifitas pengolahan data arsip, kantor camat saat ini berimplementasi manual, setiap aktifitas yang berhubungan dengan dokumentasi pihak yang berkepentingan harus mendatangi kantor kecamatan tersebut. Dari sinilah letak permasalahan yang diharuskan untuk membangun sebuah sistem dimana pihak yang berkepentingan untuk mendapatkan dokumen atau sekedar mencari informasi yang dibutuhkan cukup bisa mengakses halaman *website* yang akan dibangun dengan fitur pencarian yang menggunakan metode Algoritma Levensthein String dikarenakan metode ini dapat mengelola *string* dengan konsep pengubahan, penambahan, penghapusan, serta dapat mengukur perbedaan *string input* pencarian dengan data dokumen yang ada.

## 2  Metodologi Penelitian

Metode penelitian yang penulis gunakan dalam penelitian ini adalah metode Deskriptif. Menurut Sugiyono [6] metode deskriptif adalah suatu metode yang digunakan untuk menggambarkan atau menganalisis suatu hasil penelitian tetapi tidak digunakan untuk membuat kesimpulan yang lebih luas. Sukmadinata [7], menjelaskan penelitian deskriptif adalah suatu bentuk penelitian yang ditujukan untuk mendeskripsikan fenomena-fenomena yang ada, baik fenomena alamiah maupun fenomena buatan manusia. Fenomena itu bisa berupa bentuk, aktivitas, karakteristik, perubahan, hubungan, kesamaan, dan perbedaan antara fenomena yang satu dengan fenomena lainnya.

### 2.1 Metode Pengumpulan Data

Metode pengumpulan data yang penulis gunakan dalam menyelesaikan penelitian ini melibatkan 3 (tiga) metode, yaitu melalui : 1) observasi, 2) wawancara, 3) studi kepustakaan.

### 2.2 Metode Pengembangan Sistem

Metode Pengembangan Sistem yang digunakan pada penelitian ini adalah metode *Rational Unified Process* (RUP). RUP adalah suatu kerangka kerja proses pengembangan perangkat lunak iteratif. RUP bukanlah suatu proses tunggal dengan aturan yang konkrit, melainkan suatu kerangka proses yang dapat diadaptasi dan dimaksudkan untuk disesuaikan oleh organisasi pengembang dan tim proyek perangkat lunak yang akan memilih elemen proses sesuai dengan kebutuhannya [8].

Tahapan-tahapan dalam RUP adalah sebagai berikut : 1) *Inception*. Tahap ini memodelkan proses bisnis (*business modelling*) dan mendefinisikan kebutuhan sistem (*requirements*), 2) *Elaboration*. Tahap ini lebih difokuskan pada perencanaan arsitektur sistem. Tahap ini juga mendeteksi apakah arsitektur sistem yang diinginkan dapat dibuat atau tidak. Tahap ini lebih pada analisis dan desain sistem serta implementasi sistem yang fokus pada purwarupa sistem (*prototype*). Langkah-langkah yang dilakukan penulis pada tahap ini meliputi : perancangan *database*, alur sistem yang akan dibuat, antar muka, analisa dan desain teknis, 3) *Construction*. Tahap ini fokus pada pengembangan komponen dan fitur-fitur yang ada pada sistem. Tahap ini lebih pada implementasi dan pengujian sistem yang fokus pada implementasi perangkat lunak pada kode program. Langkah-langkah yang dilakukan meliputi, pembuatan tampilan (*layout*) dan pembuatan kode program, 4) *Transition*. Tahap ini lebih pada *deployment* atau instalasi sistem yang sudah berhasil dibuat agar dapat dimengerti oleh *user*. Langkah-langkah yang dilakukan penulis pada tahap ini meliputi, pengujian program aplikasi. Dalam metode RUP yang terbatas pada *Core Process Workflows*, perangkat lunak dikembangkan secara iterasi dengan 6 (enam) jenis kegiatan yang harus dilakukan, yaitu : 1) *Business modelling*, 2) *Requirements*, 3) *Analysiss and design*, 4) *Implementation*, 5) *Testing*, dan 6) *Deployment*.

## 3 Hasil dan Pembahasan

Proses pembuatan Sistem Informasi Pengarsipan Menggunakan Algoritma Levensthein String pada Kecamatan Seberang Ulu II dilakukan dengan metode

Fakultas Ilmu Komputer
Program Studi Sistem Informasi dan Teknik Informatika
Universitas Bina Darma

pengembangan RUP. Sistem informasi ini dikembangkan dengan bahasa PHP dengan teknik Pemrograman Berorientasi Objek. Pemodelan pada pengembangan sistem ini menggunakan UML.

### 3.1 Halaman Login dan Home

Halaman home *login* merupakan halaman awal yang menampilkan daftar menu yang biasa diakses orang banyak yang memiliki kepentingan serta sekaligus halaman *login* untuk pengguna sistem. Halaman *home* sistem ini merupakan halaman untuk pengguna sistem setelah login.

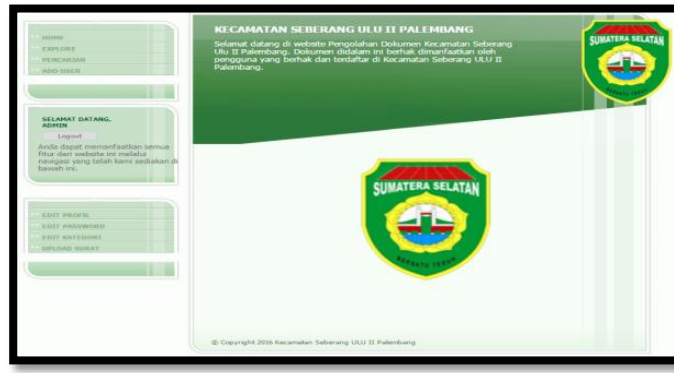

**Gambar 1.** Halaman Login dan Home.

### 3.2 Halaman Explore

Halaman *explore* merupakan halaman untuk pengelompokan atau *root* lokasi *file* berdasarkan kelompok yang telah ditentukan. Ada 4 (empat) *root*, yaitu : 1) Artikel, 2) Dokumen Surat Keluar, 3) Dokumen Surat Masuk, dan 4) Gambar.

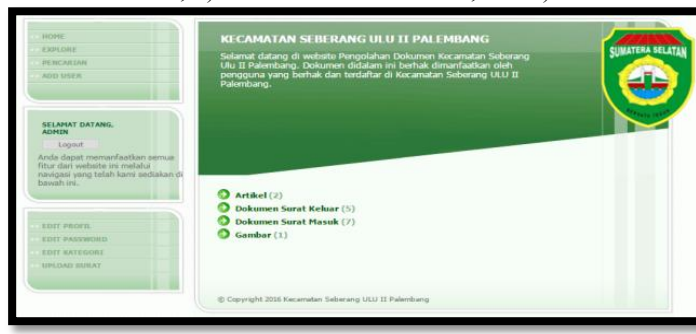

**Gambar 2.** Halaman *Explore*.

### 3.3 Halaman Pencarian

Halaman ini merupakan halaman yang berfungsi untuk mencari kumpulan *file* dengan metode pencarian *string* algoritma Levensthein. *User* dapat memasukkan *string* pada kotak *input* pencarian, kemudian klik tombol cari. Jika diperlukan bisa dilengkapi dengan memasukkan kategori yang dapat dipilih dengan menggunakan *combo box*.

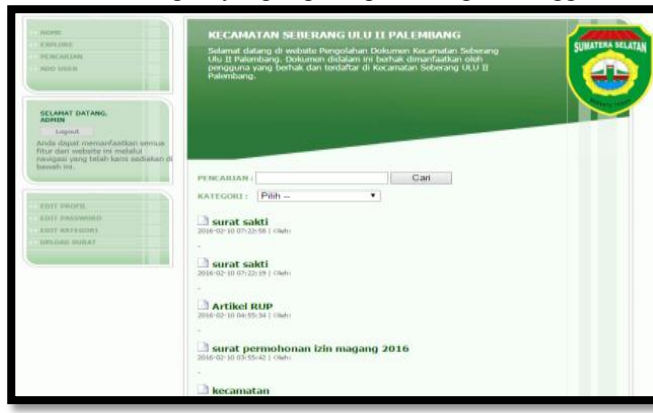

**Gambar 3.** Halaman Pencarian.

### 3.4 Halaman *Upload* Surat

Halaman *upload* surat ini digunakan untuk proses *Create, Read, Update, Delete* (CRUD) *file*. Halaman ini hanya bisa di akses untuk level *Admin*. *Fields* yang diisikan antara lain : 1) Perihal, 2) No Surat, 3) Deskripsi, 4) Kategori, dan 5) *File*. *File* yang di-*upload* berupa *file* dengan format terpopuler, yaitu *file* dengan format PDF [9].

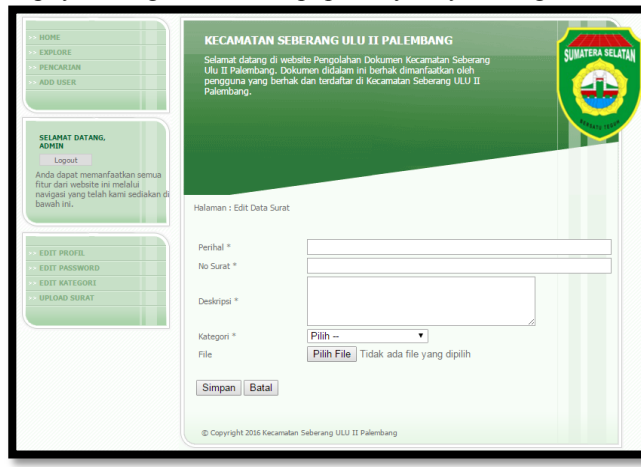

**Gambar 4.** Halaman *Upload* Surat.

11

## 4  Simpulan dan Saran

Berdasarkan pembahasan pada bab-bab sebelumnya terhadap Pengolahan Dokumen pada Kecamatan Sebrang ULU II, maka peneliti mengambil beberapa kesimpulan sebagai berikut :

1) Membangun sistem informasi pengarsipan menggunakan Algoritma Levensthein String pada Kcamatan Seberang Ulu II dengan bahasa pemrograman PHP.
2) Membantu mempermudah dalam pengolahan dokumen pengarsipan yang ada di kecamatan sebrang ulu II dengan menggunakan Algoritma Levensthein String.
3) Dengan menggunakan Algoritma Levensthein yang bersifat *Aproximate String Matching* yang dapat melakukan pencarian *strin*g khusus dengan pendekatan perkiraan.

## Daftar Pustaka

1. L. A. Abdillah and Emigawaty, "Analisis laporan tugas akhir mahasiswa Diploma I dari sudut pandang kaidah karya ilmiah dan penggunaan teknologi informasi," *Jurnal Ilmiah MATRIK,* vol. 11, pp. 19-36, April 2009.
2. E. Pratama*, et al.*, "Correspondence archival information systems in bina darma university," in *The 4th International Conference on Information Technology and Engineering Application 2015 (ICIBA2015)*, Palembang, 2015.
3. L. A. Abdillah, "Algorithms & Programming," in *Computer Science for Education*, ed. Palembang: Bina Darma University, 2013.
4. I. Mulyana*, et al.*, "Penerapan Algoritma Edit Distance untuk Pengukuran Keiripan Antar Dokumen Berbahasa Indonesia," in *Seminar Nasional Teknologi Informasi dan Komunikasi (SEMNASTIK2014)*, Palembang, 2014.
5. L. A. Abdillah, "Algorithms & Data Structures," in *Computer Science for Education*, ed. Palembang: BIna Darma University, 2016.
6. Sugiyono, *Metode Penelitian Deskriptif*. Bandung: Alfabeta, 2005.
7. Sukmadinata, *Metode Penelitian*. Bandung: Rosdakarya, 2006.
8. S. W. Ambler, "A manager's introduction to the Rational Unified Process (RUP)," *Version: December,* vol. 4, 2005.
9. L. A. Abdillah, "PDF articles metadata harvester," *Jurnal Komputer dan Informatika (JKI),* vol. 10, pp. 1-7, April 2012.